\documentclass[
aip,
jap,
final,
reprint,
preprintnumbers,
showpacs,
amsmath,
amssymb]{revtex4-1}
\usepackage{graphicx}
\usepackage{dcolumn}
\usepackage{bm}
\usepackage{textcomp}
\usepackage{revsymb}
\usepackage{wasysym}  
\usepackage{ifsym}  
\usepackage{amssymb}  
\usepackage{pifont}  
\usepackage{textcomp}
\usepackage{tipa}

\begin{document}


\title{Application of hydrogenation to low-temperature cleaning of the Si(001) surface in the processes of molecular-beam epitaxy: Investigation by STM, RHEED and HRTEM}

\author{L. V. Arapkina}
\email{arapkina@kapella.gpi.ru}
\noaffiliation
 
\author{L. A. Krylova}
\noaffiliation

\author{K.~V.~Chizh}
\noaffiliation

\author{V.~A.~Chapnin}
\noaffiliation

\author{O.~V.~Uvarov}
\noaffiliation

\author{V.~A.~Yuryev}
\altaffiliation[Also~at ]{Technopark of GPI RAS, Moscow, 119991, Russia}
\email{vyuryev@kapella.gpi.ru} 
\homepage{http://www.gpi.ru/eng/staff\_s.php?eng=1\&id=125}
\noaffiliation

\affiliation{A.\,M.\,Prokhorov General Physics Institute of the Russian Academy of Sciences, 38 Vavilov Street, Moscow, 119991, Russia}


\date{28 February 2012}

\begin{abstract}
 Structural properties of the clean Si(001) surface obtained as a result of low-temperature (470--650\textcelsius) pre-growth annealings of silicon wafers in a molecular-beam epitaxy chamber have been investigated. To decrease the cleaning temperature, a silicon surface was hydrogenated in the process of a preliminary chemical treatment in HF and NH$_4$F aqueous solutions. It has been shown that smooth surfaces composed by wide terraces separated by monoatomic steps can be obtained by dehydrogenation at the temperatures $\apprge 600${\textcelsius}, whereas clean surfaces obtained at the temperatures $< 600${\textcelsius} are rough. It has been found that there exists a dependence of structural properties of clean surfaces on the temperature of hydrogen thermal desorption  and the process of the preliminary chemical treatment. The frequency of detachment/attachment of Si dimers from/to the steps and effect of the Ehrlich-Schwoebel barrier on ad-dimer migration across steps have been found to be the most probable factors determining a degree of the resultant surface roughness.
 \end{abstract}



\maketitle

\section{Introduction }

Reduction of temperatures of technological processes is one of important requirements of silicon VLSI manufacturing.\cite{VCIAN2011}  Presently, removal of a SiO$_2$ protection layer, formed in the process of wet chemical etching, with the use of thermal treatments at temperatures of $\apprge 800${\textcelsius}  is a main method of pre-growth cleaning of Si surfaces before  molecular-beam epitaxy (MBE) and the RCA method\cite{RCA,cleaning_handbook} is most frequently used for formation of a thin ($\sim 1$~nm thick) film  of silicon dioxide.\cite{Shiraki} Application of  Si(001) surface hydrogenation instead of its oxidation at the final stage of wet etching  may enable the reduction of the surface cleaning temperature down to 500--650{\textcelsius}.\cite{Point_defects_to_huts,thermal_desorption} In this article, an issue of possibility of reduction of Si(001) wafer cleaning temperature before MBE, directly in the ultra-high vacuum (UHV) deposition chamber, is investigated by means of three complimentary techniques: combination of the scanning tunneling  microscopy (STM) and the reflected high-energy electron diffraction (RHEED) allows us to reveal and study the Si(001) surface morphology whereas the  high resolution transmission electron microscopy (HRTEM) is used for examination of epitaxial layers and interfaces formed as a result of Si deposition on the dehydrogenated Si(001) surfaces. 

Studies of properties of hydrogenated Si(001) and Si(111) surfaces have been carried out for a long time. Initially these surfaces were explored using diffraction methods, the method of thermal desorption and the surface IR spectroscopy.\cite{cleaning_handbook,thermal_desorption,ELS_Si(111),H-termination_Si(111), IR_Si_oxidation,In_situ-etching,H-terminated_morphology,Hydrogen-terminated_Si(111)-Si(100)} Results of these researches enabled the determination of types of hydrides which form on surfaces of differently oriented Si wafers depending on composition of etchants as well as the temperature range in which their desorption is possible.

With the development of the STM techniques, it has become possible to conduct researches of structural properties of hydrogenated and clean surfaces at a qualitatively new level.\cite{Hydrogen_interaction,H-saturated_Si(100),STM_2x1_to_3x1} In the vast majority of the works, the hydrogenation of the $(2\times 1)$-reconstructed Si(001) surface purified of the silicon dioxide film was carried out under UHV using a source of atomic hydrogen. A structure of surfaces obtained in such a way differs significantly from the structure of surfaces forming as a result of wet etching without hydrogenation. Annealing at high temperatures for SiO$_2$ removal  results in formation of surfaces with wide terraces and monoatomic steps and facilitates the desorption of foreign atoms.\cite{Empty_states_Si(100)-(2x1),Mosaic_Si(111)-(7x7)} STM studies have shown that Si(001) surfaces hydrogenated as a result of wet etching are composed by narrow terraces with monoatomic steps.\cite{STM_hydrogen-terminated_Si(001),Si(001)_HF_cleaning,Si(001)_wet_cleaning_STM,Aqueous_Etching_Si(100)} Impossibility of desorption of some impurities, e.g., carbon, by a subsequent low-temperature treatment is a disadvantage of the surface cleaning method based on its hydrogenation during wet etching.

Short dipping of a Si wafer in an aqueous solution of the hydrofluoric acid with pH varying from 2 to 7 is the most well-known method of the surface hydrogenation. A number of Si hydrides, the most prevalent of which are silicon monohydride and dihydride, form  as a result of reaction of hydrogen with the Si surface.\cite{H-saturated_Si(100),Aqueous_Etching_Si(100),H-termination_Si(001)_wet,Si(100)H} A fraction of dihydride on the surface rises  as pH of the HF solutions increases.\cite{IR_Si_oxidation,In_situ-etching} In some modes of Si(001) etching, formation of the \{111\} facets is observed, on which every Si atom located on the (001) plane is bonded to two hydrogen atoms (dihydride) whereas each atom situated on the (111) plane is bonded to a single atom of hydrogen (monohydride).\cite{H-termination_Si(111),H-terminated_morphology}  The temperature of thermal desorption of hydrogen atoms depends on a type of the hydride. According to Ref.\,\onlinecite{thermal_desorption}, a characteristic temperature range of the monohydride thermal desorption is from 540 to 620\textcelsius, that of the dihydride desorption is from 440 to 540\textcelsius.

In the current article, the data of exploration by means of STM and RHEED of the structure of Si(001) surfaces obtained as a result of the low-temperature dehydrogenation of silicon substrates after wet chemical etching with surface termination by hydrogen atoms are presented. Results of investigation of surface structure of Si epitaxial layers grown on such surfaces as well as data of HRTEM studies of Si-epilayer/Si-substrate interfaces  are also set forth.

  \section{Experimental Techniques and Sample Preparation Procedures}

The investigations were carried out using the Riber EVA32 MBE vessel equipped with the Staib Instruments RH20 RHEED tool and coupled with the GPI-300 UHV STM.\cite{VCIAN2011,stm-rheed-EMRS,classification,*CMOS-compatible-EMRS}   8$\times$8\,mm$^2$ samples were cut from commercial CZ Si:B wafers (the orientation was (100), $\rho=12\,\Omega$\,cm).

During the preliminary chemical treatment, at final phase of RCA wet etching, the samples were processed in various aqueous solutions on the basis of the hydrofluoric acid. The following etchants were used: (1) $5\%$ HF solution, pH~=~2; (2) buffered NH$_4$F\,+\,HF solution, pH~=~4; (3) concentrated NH$_4$F solution, pH~=~7. The chemical processing was completed by sample rinsing in deionized water for 1 min.

The thermal treatment in UHV was performed in two stages: (i) preliminary annealing lasted 2 hours at 300{\textcelsius}---this process completed the preliminary treatment of samples; (ii) final annealing was carried out at higher temperature in the range from 470 to 650{\textcelsius}. The duration of the second stage of annealing was determined by a form of RHEED patterns. 

The process temperature was monitored by the IS12-Si infrared pyrometer.  The pressure in the MBE chamber did not exceed $8\times 10^{-10}$\,Torr during annealing. The rate of sample heating from 300{\textcelsius} to the preset temperature of final annealing was $\sim$\,0,24\,{\textcelsius}/c; the sample cooling rate was $\sim$\,0,4\,{\textcelsius}/c. 

The STM images were taken in the constant current mode at room temperature.

Si epitaxial films were deposited on the clean Si(001) surface by UHV MBE. Conditions in the MBE chamber during Si deposition were similar to those during Ge/Si(001) MBE described  in Refs.\,\onlinecite{VCIAN2011,classification,*CMOS-compatible-EMRS}. The rate of Si deposition was $\sim$\,0.25\,\AA/c, the sample cooling rate after growth was $\sim$\,0.4\,{\textcelsius}/c.

The Carl Zeiss Libra-200 FE HR HRTEM was used for investigation of structural properties of Si epitaxial layers and their interfaces with Si substrates. 

Additional experimental details and description of the experimental instruments can be found in Ref.\,\onlinecite{VCIAN2011}.


\section{Exploration by STM and RHEED}

\subsection{Hydrogenated Si(001)}

\begin{figure*}[t]
\includegraphics[scale=1.35]{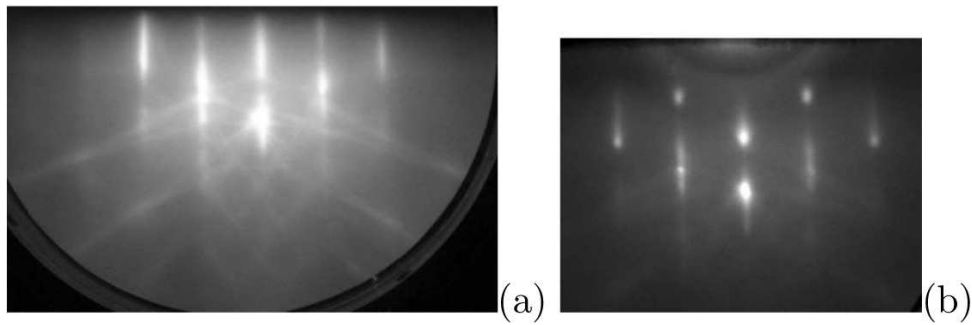} 
\caption{\label{fig:RHEED}
RHEED patterns of the Si:H surfaces obtained as a result of treatment in solutions of varying acidity (azimuth is [110], $E = 10$~keV'): (a)~pH~=~2; (b)~pH~=~7.
}
\end{figure*}

A study of silicon surfaces terminated by hydrogen (Si:H)  carried out  by means of STM and RHEED has shown that their structure depends on a kind of the preliminary chemical treatment used for hydrogenation. If solutions containing NH$_4$F are used Si:H surfaces are more rough in comparison with those prepared in solutions of dilute HF; the higher the concentration of ammonium fluoride, the rougher the surface forms. Surface roughening is a consequence of formation of surface islands composed by terraces and monoatomic steps. Height and area of the islands are determined by conditions of the preliminary chemical treatment. Difference in surface roughness may be qualitatively estimated from the RHEED patterns. Irrespective of a kind of the solution used for the surface termination, RHEED patterns always correspond to the unreconstructed $1\times 1$ surface (Fig.\,\ref{fig:RHEED}).

Diffraction patterns of the samples prepared in solutions with   pH~=~2 consist of broad streaks of main reflexes  (Fig.\,\ref{fig:RHEED}a), whereas beams of main reflexes have shapes corresponding to well developed roughness on the monoatomic scale (3D reflexes) for the samples processed in solutions with  pH~=~7 (Fig.\,\ref{fig:RHEED}b).

Effect of NH$_4$F concentration on the roughness of resultant surfaces consists in selective etching of surfaces and appearance of the \{111\} facets during sample treatment in solutions containing ammonium fluoride. However, it is demonstrated in the literature  that formation of such facets is suppressed if intense release of hydrogen bubbles is eliminated.\cite{same_etchant} STM study of the Si:H surfaces has shown a slight increase in the spread of height   of adjacent steps on the surfaces with the growth of etchant pH---it is within the height of two monoatomic steps. We have failed to recognize a type of the Si:H surface reconstruction: it is difficult to be determined because the hydrides may fill the surface irregularly. Another circumstance complicating the recognition of a type of the Si:H surface reconstruction is that hydrogen atoms situated close to one another but bonded to different silicon atoms ``deviate'' from an ``ideal'' position asymmetrically and non-uniformly around the surface. This results in difference of the local density of states (LDOS) in the vicinity of a Si atom from LDOS in the vicinity of another Si atom, therefore atomic resolution cannot be reached in the STM images of such surfaces.\cite{H-termination_Si(001)_wet,Si(100)H}

\subsection{Effect of surface dehydrogenation}

Exploration of the Si(001) surface cleaned of hydrogen atoms by thermal desorption allows one to mark out two characteristic ranges of the treatment temperature---below and above 600{\textcelsius}---in which physical processes result in formation of different surface structures. 

\subsubsection{Annealing at T $\mathbf{\apprge}$ 600\textcelsius }

\begin{figure*}[t]
\includegraphics[scale=1]{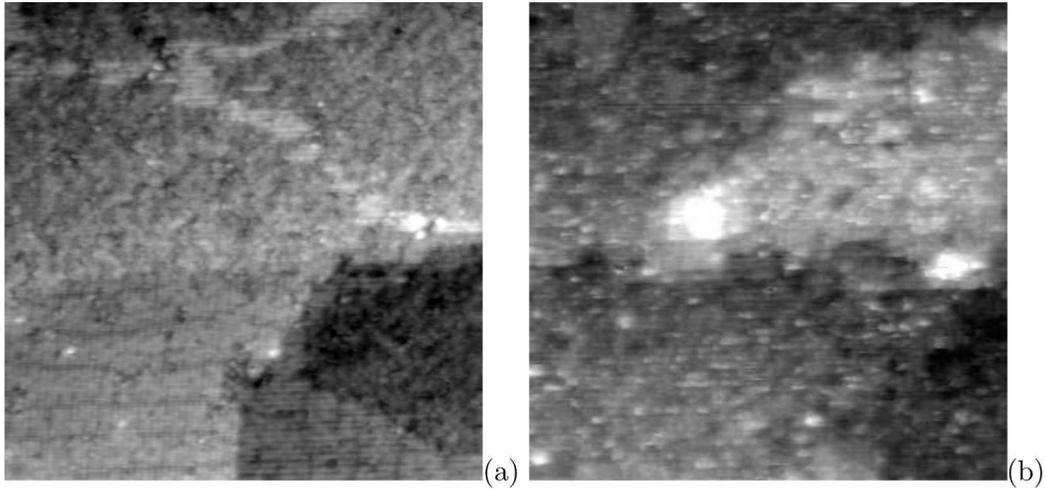}
\caption{\label{fig:STM}
STM images of the  Si(001) surfaces subjected to preliminary chemical treatments in solutions of varying acidity and dehydrogenated as a result of  annealing at  $T=650$\textcelsius: (a)~pH~=~2, annealing for 8 min, $57\times 57$ nm; (b)~pH~=~7, annealing for 5 min, $40\times 40$ nm.
}
\end{figure*}

If the cleaning temperatures are higher than 600{\textcelsius} enhancement of NH$_4$F content in solutions during the preliminary etching  causes an increase of the resultant surface roughness; with the growth of the NH$_4$F concentration longer annealing is required for obtaining a smoother surface. Smooth surfaces with wide terraces and monoatomic steps form as a result of processing in solutions with pH~=~2 followed by annealing at the temperatures above 600{\textcelsius} or as a result of etching in solutions with pH~$>$~2 and annealing at the temperatures $\apprge$\,650{\textcelsius}. Fig.\,\ref{fig:STM} shows STM images of the clean Si(001) surface subjected to difference chemical treatments and annealed at 650{\textcelsius}.

The  $c(4\times 4)$ structure arises on the surface due to annealing at 650{\textcelsius}. Previously this structure was identified by us as $2\times 2$,\cite{phase_transition} however further survey have shown this structure to be nothing else but the $\beta$-modification of the $c(4\times 4)$ reconstruction.\cite{Uhrberg,C-coverage} One can find a detailed description of the Si(001)-$c(4\times 4)$ structure in Ref.\,\onlinecite{VCIAN2011}.

Reflexes corresponding to the $c(4\times 4)$ surface structure are much weaker in the RHEED patterns of the samples processed in the solutions with pH~=~7 than in the diffraction patterns of the samples treated in the solutions with pH~=~2. STM images show that $c(4\times 4)$ reconstructed surfaces are always composed by wide terraces and monoatomic steps. A relation between processes of the $c(4\times 4)$ structure formation and obtaining of smooth surfaces is difficult to be ascertained because it is hard to establish if there is a contribution of carbon to these processes: according to existing notions it can take part in formation of the $c(4\times 4)$ structure.\cite{Miki,Kim-Kim} It is obvious however that Si atoms (which are known to be paired in ad-dimers) can freely migrate along the surface and  the $c(4\times 4)$ structure forms on the $2\times 1$ reconstructed surface. Evidence of this fact has been found in data on evolution of the diffraction patterns during heat treatments: the $c(4\times 4)$ structure appears during sample exposure at the temperatures from 610 to 650{\textcelsius} but it does not arise in course of rapid heating of samples in the interval from 300 to 800{\textcelsius}.

\begin{figure*}[t]
\includegraphics[scale=1]{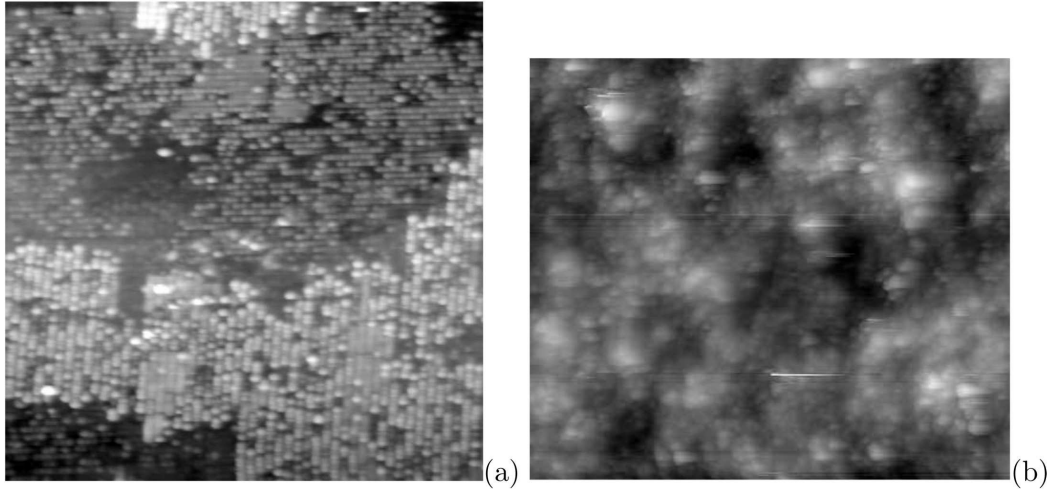}
\caption{\label{fig:610} 
STM images of the  Si(001) surfaces subjected to preliminary chemical treatments in solutions of varying acidity and dehydrogenated by  annealing at   $T=610${\textcelsius} for 10 min: (a)~pH~=~2,  $41\times 41$ nm; (b)~pH~=~7, $87\times 76$ nm.
}
\end{figure*}

Differences in the structure of the formed clean surfaces are also observed in the samples dehydrogenated by annealing at 610{\textcelsius} (Fig.\,\ref{fig:610}). The $c(4\times 4)$ structure is observed on the samples treated in the solutions with  pH~=~2 and annealed for 10 min. Further prolongation of annealing may cause occurrence of SiC islands on the surface (although we have never met SiC on our samples this circumstance should be taken into account). When hydrogenated in the solutions with pH~=~7, the clean surface becomes rougher, it is composed by islands formed by short narrow terraces and monoatomic steps. Diffraction patterns correspond to the $2\times 1$ reconstruction; they are characterized by weaker 3D reflexes on the main beams than those shown in Fig.\,\ref{fig:RHEED}b.

\subsubsection{Annealing at T $\mathbf{<}$ 600\textcelsius }

\begin{figure*}[t]
\includegraphics[scale=1]{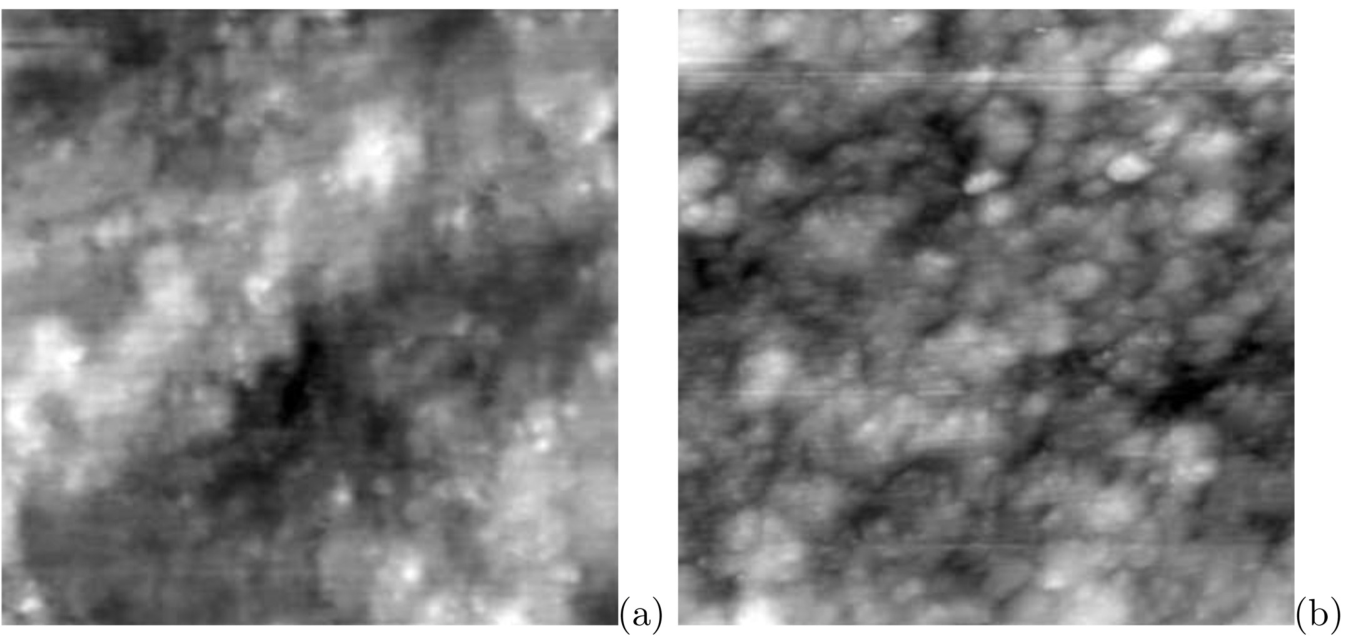} 
\caption{\label{fig:530-550}
STM images of the  Si(001) surfaces subjected to preliminary chemical treatments in solutions of varying acidity and dehydrogenated at lowered temperatures: (a)~pH~=~2, $T=530$\textcelsius, $41\times 41$ nm; (b)~pH~=~7, $T=550$\textcelsius, $60\times 60$ nm.
}
\end{figure*}

Annealings at the temperatures below 600{\textcelsius} result in formation of clean surfaces with the roughness on monoatomic scale for any type of the preliminary chemical treatment. However the structures of the resultant surfaces are different for different chemical treatments. Hydrogenation in solutions with greater pH results in formation of rougher surfaces. STM images of samples annealed at the temperatures below 600{\textcelsius} after different chemical treatments  are presented in Fig.\,\ref{fig:530-550}.

The minimum temperature at which surface cleaning takes place is 470{\textcelsius} (annealing duration is 35 min) for the samples treated in the solutions with pH~=~2 and 550{\textcelsius} (annealing duration is also 35 min) for the samples processed in the solutions with pH~=~7. Diffraction patterns correspond to the $2\times 1$ structure for the former process; for the latter processing they resemble the $1\times 1$ patterns of the original surface terminated by hydrogen atoms but with narrower stripes and less pronounced 3D reflexes on the main beams  (the {\textonehalf} reflexes are always weak when the processing temperature is less than 600{\textcelsius} but in the case of etching at pH~=~7 they are usually invisible to the eye and can be registered only with a camera at a long exposure). 

\subsubsection{Dynamics of RHEED patterns during sample heating}

A study of evolution of the diffraction patterns during sample heating from 300 to 800{\textcelsius} has shown that original patterns correspond to the $1\times 1$ structure, i.\,e., they are composed only by the main beams. 3D reflexes are observed on the main beams; the brightness of the 3D reflexes depends on the preliminary chemical treatment. The most intense 3D reflexes are observed after the treatment in the solutions containing NH$_4$F. A temperature at which the {\textonehalf} beams (which correspond to  $2\times 1$ or $c(4\times 4)$ structure) arise in the [110] azimuth in electron diffraction patterns depends on the initial chemical treatment. If the samples are etched  in dilute HF the {\textonehalf} reflexes appear in the range from 420 to 470{\textcelsius}; they arise in the range from 560 to 660{\textcelsius} or do not appear at all  if the samples are hydrogenated in the solutions with  pH~=~4 or 7. The cause of the difference in the temperatures at which the {\textonehalf} reflexes appear lies in the strong dependence of the original Si:H surface roughness on  the composition of the solutions used for the surface termination.

It can be concluded  that RHEED can be applied as a method for monitoring of hydrogen thermal desorption from the Si:H surfaces only in the case if solutions are used for the Si(001) surface hydrogenation, the treatment in which results in formation of the Si:H surfaces with low roughness. Strong roughness of a clean surface interferes with formation of sufficient $2\times 1$ reconstructed areas and application of RHEED for monitoring of hydrogen desorption  becomes impossible.

\begin{figure*}[th]
\includegraphics[scale=1]{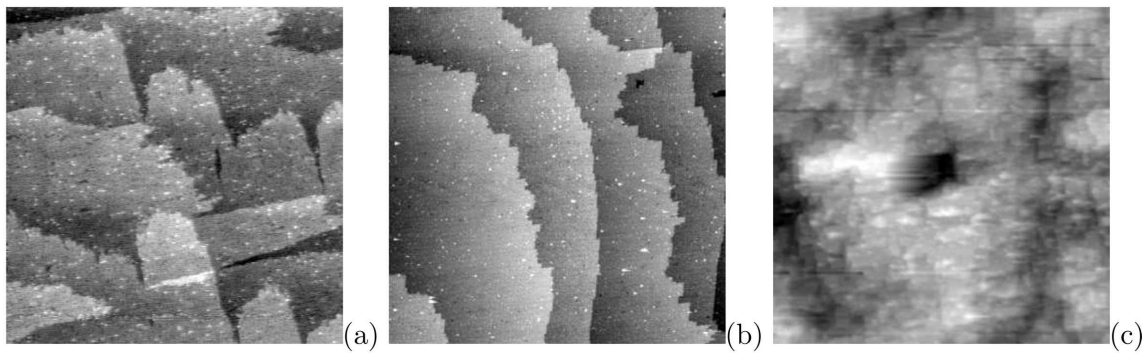} 
\caption{\label{fig:epi-Si}
STM images of 50~nm thick Si/Si(001) films  grown by MBE at different temperatures: (a)~610\textcelsius, $41\times 41$ nm; (b)~550\textcelsius, $60\times 60$ nm; (c)~470\textcelsius, $60\times 60$ nm.
}
\end{figure*}

\begin{figure*}[ht]
\includegraphics[scale=1]{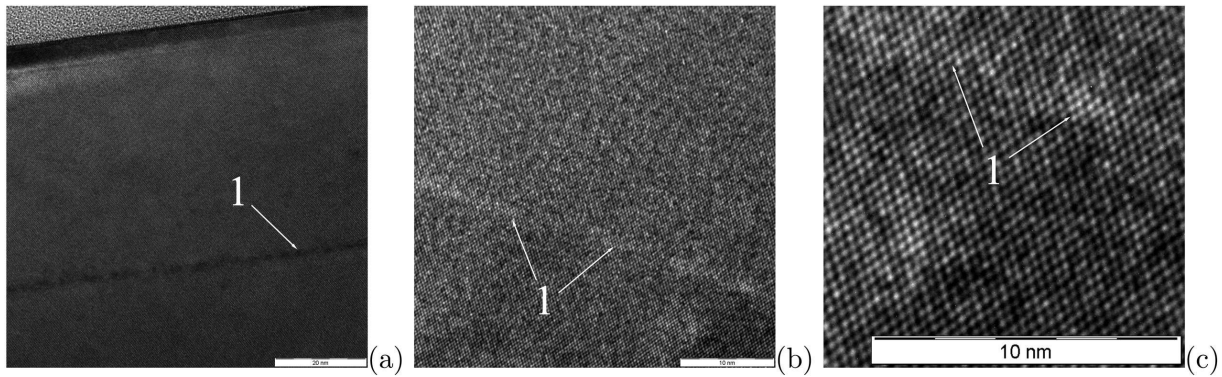} 
\caption{\label{fig:TEM_610C}
HRTEM micrographs of a 50~nm thick Si epitaxial film  grown at 610{\textcelsius} on Si(001) after its preliminary chemical treatment in the solution with pH~=~2 followed by dehydrogenation at {610\textcelsius} for 10~min; the figure 1 indicates an epi-Si/substrate interface region; a scale bar corresponds to 20~(a) or 10~(b),(c)~nm; panel (c) is magnified from  (b); a spotty triangle in the left upper corner of the image (a) is an epoxy.
}
\end{figure*}

\subsubsection{Brief discussion of annealing temperature effect on surface structure}

It is evident from the above that a degree of roughness of clean Si(001) surfaces with monoatomic steps depends on the temperature of hydrogen thermal desorption. Thermal treatments at a temperature above 600{\textcelsius} enable obtaining of surfaces with wide terraces and monoatomic steps. Heat treatments at a temperature below 600{\textcelsius} do not result in formation of smooth surfaces, the surface roughness decreases insignificantly during annealing. This effect is likely caused by the change in conditions of surface migration of Si atoms. At the temperatures higher than 600{\textcelsius} islands composed by monoatomic steps and terraces---emerged as a result of chemical treatments---``dissolve'' with emission of Si ad-atoms (ad-dimers) which cannot desorb from the surface, migrate, build in steps or embed in the surface being captured by its defects. Thus, initially rough surface smoothens.  

At the temperatures lower than 600{\textcelsius} a surface also slightly smoothens but wide terraces do not arise. According to Refs.\,\onlinecite{thermal_desorption,In_situ-etching,H-terminated_morphology,Goldfarb_JVST-A} incomplete hydrogen thermal desorption should not affect the smoothing process as the temperature interval at which annealings were carried out in this work and the sample heating rate should ensure complete removal of hydrogen atoms from the surface for the period of time much shorter than  duration of the  annealings applied. The described effect may result from slowdown of Si dimer migration on the surface with lowering of the temperature  to $\sim$\,600{\textcelsius}. In addition to reduction of the ad-dimer mobility (hop frequency), the relief formation starts to be affected by decreasing rate of detachment  of  Si dimers from and their attachment to surface steps.\cite{Variable-temperature_STM,Si(001)_step_dynamics} 

It is known that the attachment/detachment of Si at the monatomic steps goes in units of two dimers.\cite{surface-dynamics_Si(001)} The latter has to do with the fact that a kink can terminate on top or in between two dimer rows: one of these configurations is energetically more favorable than the other. Below 600{\textcelsius} the concentration of  mobile dimers thermally detached from the steps is lower than at temperature above 600{\textcelsius}. In order to have significant mass transport (step flow) the dimer that descends from a step should meet another mobile dimer in order to be ``built in'' at the kink sites (two dimers are required in order to result in step flow). At high temperatures the concentration of mobile dimers is higher than at low temperatures therefore a probability that a kink can propagate will increase at higher temperatures leading to more mass transport and better equilibrated surfaces. At the same time lower density of ad-dimers explains higher roughness of surfaces at low temperatures.

Additionally, an influence of the Ehrlich-Schwoebel (ES) barrier\cite{Schwoebel,Ehrlich-Schwoebel}---a potential barrier on terrace edges---becomes stronger at low temperatures due to ad-dimer kinetic energy reduction; as a result, most of Si atoms detaching from the upper step cannot leave the terrace to move to the next lower one; as a consequence an effective smoothing of the surface relief and formation of wide terraces do not happen. At high temperatures the ad-dimer kinetic energy appreciably exceeds the ES barrier height and the effect of the barrier on dimer migration becomes negligible, so the steps at these temperatures are permeable for adatoms.\cite{permeable_step} This favours the development of smooth surfaces with wide terraces.


\subsection{TEM study of epi-Si/Si-substrate interfaces}

\begin{figure*}[th]
\includegraphics[scale=1]{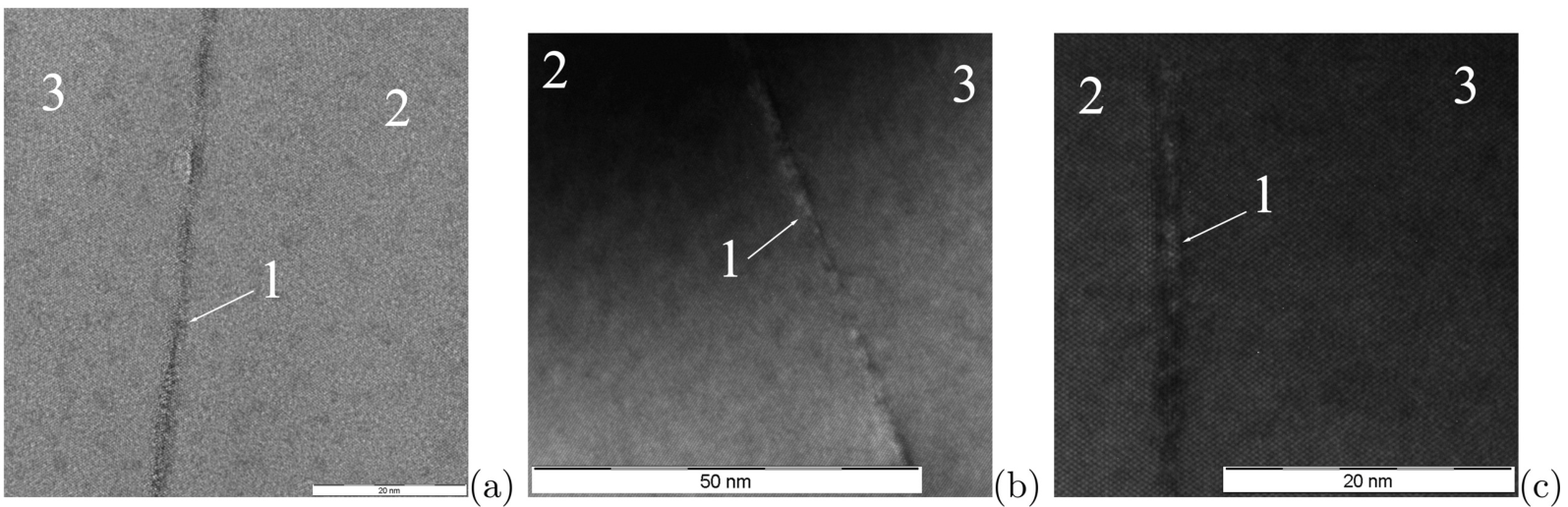} 
\caption{\label{fig:TEM_550C}
HRTEM micrographs of a 50~nm thick Si epitaxial film  grown at 550{\textcelsius} on Si(001) after its preliminary chemical treatment in the solution with pH~=~2 followed by annealing  at {550\textcelsius} for 30~min; the figure 1 indicates an epi-Si/substrate interface region, 2 shows a Si substrate and 3 denotes an epitaxial layer;  a scale bar corresponds to 20~(a),(c) or 50~(b)~nm.
}
\end{figure*}

\begin{figure*}[th]
\includegraphics[scale=1]{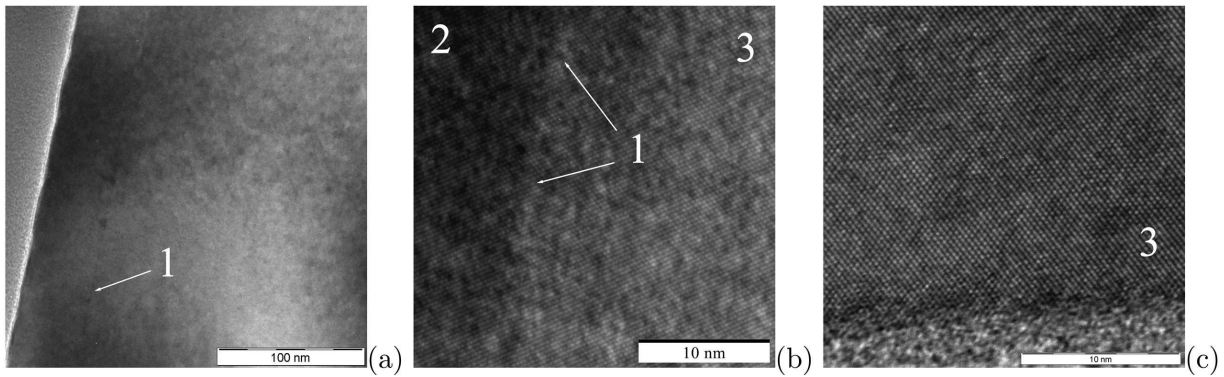}
\caption{\label{fig:TEM_470C}
HRTEM micrographs of a 50~nm thick Si epitaxial film  grown at 470{\textcelsius} on Si(001) after its preliminary chemical treatment in the solution with pH~=~2 followed by annealing  at {470\textcelsius} for 18~min; the numerical 1 indicates an epi-Si/substrate interface region,  2 denotes a Si substrate and 3 marks an epitaxial layer;  a scale bar corresponds to 100~(a) or 10~(b),(c)~nm; epoxy is seen in left upper corner of the image (a) and lower part of the image (c).
}
\end{figure*}

Usually, Si epitaxial layers with wide terraces and monoatomic steps on the surface are obtained by MBE on Si(001) if deposited in the temperature range from 500 to 600{\textcelsius}.\cite{Analysis_growth_75-mm,Kinetic_growth_instability_Si(001)} Fig.\,\ref{fig:epi-Si} presents STM images of surfaces of Si epitaxial layers grown  at different deposition temperatures on the Si(001) surface cleaned of hydrogen atoms. The obtained surfaces  are seen to be composed by terrases separated by monoatomic steps. By lowering the growth temperature, reduction of the width of the terraces is observed but their structure   is different from that of the dehydrogenated surfaces. This allows us to assume that effect of mobility (hop rate) reduction of migrating Si dimers on smoothing of dehydrogenated surfaces in this temperature interval is not as significant as an impact of two other factors (reduction of detachment/attachment rate of Si dimers from/to  steps and Ehrlich-Schwoebel potential barrier) as a result of which appreciable smoothing of the dehydrogenated rough Si(001) surfaces does not happen during annealing at the temperatures below 600{\textcelsius} in the absence of an external source of Si atoms.

Change of surface structural properties   at the temperatures of $\sim$\,600{\textcelsius} was reported by us previously in the articles on the study of the clean Si(001) surface obtained after removal of the  SiO$_2$ protection film by short high-temperature annealings.\cite{stm-rheed-EMRS,our_Si(001)_en,phase_transition} During sample cooling from the annealing temperature of  $\sim$\,925{\textcelsius},  a reversible phase transition occurs in the range from 550 to 600{\textcelsius} which is manifested as a change of the surface reconstruction: $2\times 1 \longleftrightarrow c(8\times 8)$. This transition was explained by slowdown of Si ad-dimer migration on the surface with the decrease in the sample temperature. Changes in the surface structural properties in this temperature interval were also observed in the study of Ge quantum dot nucleation on the Ge/Si(001) wetting layer and hut-cluster array formation.\cite{VCIAN2011,Nanophysics-2011_Ge,Nucleation_high-temperatures,classification,*CMOS-compatible-EMRS} The temperature of $\sim$\,600{\textcelsius} is a frontier between modes of preferential growth of huts and domes on the Ge/Si(001) wetting layer that is likely connected with changes in Ge ad-dimer migration process on the wetting layer surface at this temperature which also may be connected with the effect of the ES barrier on Ge dimer migration on the wetting layer  at low temperatures and absence of such effect at high temperatures.\cite{Nucleation_high-temperatures}

A study   of Si substrates subjected to the described above treatments  with Si epitaxial films grown on their surfaces (Fig.\,\ref{fig:epi-Si}) performed by means of high resolution TEM allowed us to evaluate the quality of interfaces between substrates and epilayers. We have succeeded to visualize the interfaces  between epilayers grown at different temperatures and surfaces exposed to different chemical treatments and various annealings and found interfaces formed on surfaces chemically treated  at pH~=~2 to be perfect.

Fig.\,\ref{fig:TEM_610C} demonstrates HRTEM images of a 50 nm thick epilayer grown at  610{\textcelsius} on the surface hydrogenated in the solution with  pH~=~2 and exposed to annealing at 610{\textcelsius} for 10 minutes for the hydrogen removal (see Fig.\,\ref{fig:610}a). An interface region is indicated by arrows and marked by the figure 1. It is visualized as a dark or light line depending on the imaging conditions. Fig.\,\ref{fig:TEM_610C}b and especially panel (c), which is magnified from (b), demonstrate a perfect atomic structure of the near-interface region and absence of extended structural defects. The observed contrast supposedly results from  stress arising around the interface because of the presence of residual hydrogen or probably minor amount of carbon (some carbon may come on the surface from chemical reagents or MBE vessel atmosphere---CO is the main after hydrogen residual gas in this chamber, a minor amount of hydrocarbons is also present in the chamber---or even from the bulk of a Czochralsky grown substrate); a slight excess of vacancies also might explain the contrast. However, a good epitaxial interface is formed as a result of this process.

A lower temperature of growth and dehydrogenation  does not result in significant worsening of the interface. Fig.\,\ref{fig:TEM_550C} shows a film deposited at  550{\textcelsius} on the surface treated at 550{\textcelsius} for 30 minutes after hydrogenation in the etchant with pH~=~2. The epi-Si/substrate interface in this sample is also epitaxial; its structure is also perfect, it does not contain extended defects. However it looks more pronounced (Fig.\,\ref{fig:TEM_550C}c) than the interface obtained at the higher temperature, maybe because of rougher initial surface and higher concentration of point defects and such residual impurities as hydrogen or carbon.

We succeeded to lower the temperature of epitaxial growth and surface annealing to 470{\textcelsius} and obtain a perfect defectless interface. Fig.\,\ref{fig:TEM_470C} represents micrographs of an epilayer grown at this temperature on the surface treated at 470{\textcelsius} for 18 minutes after etching at pH~=~2. A perfect interface (Fig.\,\ref{fig:TEM_470C}b) and epitaxial layer (Fig.\,\ref{fig:TEM_470C}c) confirm our assumption that preparation of clean Si(001) surfaces suitable for further growth of perfect epitaxial layers is possible at temperatures as low  as 470{\textcelsius}.\cite{VCIAN2011}

\section{Conclusion}

We can conclude now that a method of hydrogenation of a Si(001) surface  is applicable for low-temperature cleaning of Si substrates in UHV MBE vessels. Varying a type of the preliminary chemical treatment and a temperature of the dehydrogenation process, clean surfaces with monoatomic roughness but different width of terraces separated by monoatomic steps can be obtained. Low-temperature cleaning of Si(001) surfaces may find numerous applications. In particular, Si surface cleaning by annealing at a temperature lower than 600{\textcelsius} may be used in processes requiring preservation of a created in advance relief on the Si(001) surface.

Perfect Si epitaxial films and epi-Si/Si-substrate interfaces can be formed at the temperatures as low as 470{\textcelsius} on clean Si(001) surfaces prepared by  hydrogenation in HF aqueous solution with pH~=~2 followed by  low-temperature dehydrogenation at the temperature of 470{\textcelsius}. This result makes it possible to form low-temperature buffer layers in the process of MBE growth of  Ge/Si(001) heterostructures compatible with the standard CMOS cycle of microelectronic  device fabrication.


\begin{acknowledgments}

We appreciate financial and technological support of this research: 
 it was financed by the Ministry of Education and Science of the Russian Federation under the State Contracts No.\,$\Pi2367$, No.\,14.740.11.0069 and  No.\,16.513.11.3046;
the study was performed using the equipment of the Center of Collective Use of Scientific Equipment of GPI RAS.

\end{acknowledgments}



%

\end{document}